\documentclass[12pt]{article}
\usepackage{hyperref}
\usepackage{showkeys}
\usepackage{amsmath}
\usepackage{amssymb}
\renewcommand{\title}[1]{%
    \bigskip%
    \begin{center}%
    \Large\bf #1%
    \end{center}%
    \vskip .2in}

\renewcommand{\author}[1]{%
    {\begin{center}
    #1
    \end{center}}}
\newcommand{\address}[1]{\vspace{-1.7em}\vspace{0pt}
    {\begin{center}
    \it #1
    \end{center}}}

\begin{document}


\title{General algorithm for non-relativistic diffeomorphism invariance}

\author
{
Rabin Banerjee  $\,^{\rm a,b}$,
Arpita Mitra    $\,^{\rm a, e}$,
Pradip Mukherjee $\,^{\rm c,d}$}
\address{$^{\rm a}$S. N. Bose National Centre 
for Basic Sciences, JD Block, Sector III, Salt Lake City, Kolkata -700 098, India }

\address{$^{\rm c}$Department of Physics, Barasat Government College,\\10 KNC Rd., Barasat, kolkata 700 124, India

 }

\address{$^{\rm b}$\tt rabin@bose.res.in}
\address{$^{\rm e}$\tt arpita12t@bose.res.in}
\address{$^{\rm d}$\tt mukhpradip@gmail.com}

\begin{abstract} 
 An algorithmic approach towards the formulation of non-relativistic diffeomorphism invariance has been developed which involves both matter and gauge fields. A step by step procedure has been provided which can accommodate all types of (abelian) gauge interaction. The algorithm is  applied to the problem of a two dimensional electron moving under an external field and also under the Chern-Simons dynamics.
\end{abstract}
\section{Introduction}

      The formulation of non-relativistic theories on a space-time manifold dates back to the works of Elie Cartan \cite{Cartan}, the corresponding manifold being named as Newton-Cartan space-time. Subsequently, investigations of different aspects of Newton-Cartan space-time have been performed by many stalwarts \cite{Havas} - \cite{MALA}. The main thrust of these works was to interpret Newtonian gravity as a space-time phenomenon. 
      
      Recently, non-relativistic theories and their associated symmetries based on non-relativistic diffeomorphism invariance have received renewed attention owing to applications in mesoscopic physics, especially, the theory of fractional quantum Hall effect. Here the first objective is to obtain a generally covariant theory in the non-relativistic perspective i.e. which has Galilean invariance in the flat (euclidean) space and universal time. Naturally the problem of non-relativistic diffeomorphism is being pursued with renewed vigor in the current literature. Consequently, various approaches are gradually emerging \cite{SW}-\cite{KjK}.

         One approach introduces spatial diffeomorphism by assuming definite transformation properties of the fields by inspection so that the theory at hand is generally covariant \cite{SW} in three dimensional space. The physical theory is (2 + 1) dimensional non-relativistic electrodynamics where the gauge field is either an external field \cite{SW} or dynamically  included in the system \cite{HS, S}. For an external gauge field which transforms as a vector under general coordinate transformation only time-independent coordinate transformations are allowed. In this context time dependent transformations may be accommodated but the gauge field no longer satisfies the usual transformations and the transition to flat space is not clear. On the other hand when  the gauge field dynamics is given by the Chern-Simons (C-S) term the general covariance is lost \cite{HS, S} and can be regained only by including additional fields.
  
    An algebraic approach to the problem has been advanced in \cite{BP}. This is 
based on a contraction of the Poincare gauge group to centrally extended Galilean group. The well known procedure of obtaining Riemann-Cartan spacetime from gauging the Poincare algebra is used to obtain the Newton-Cartan spacetime. 
This is thus an algebraic approach which still leaves the question unanswered -- how to systematically build a diffeomorhism invariant field theory that corresponds to a theory invariant under the full (extended) set of Galilean transformations in the flat limit? A field theoretic approach is required therefore. 

In a recent paper \cite {BMM1} we have provided a systematic method of constructing a non-relativistic diffeomorphism invariant field theory that has the appropriate flat limit. This procedure is inspired by the celebrated Poincare Gauge Theory (PGT) [\cite{U} - \cite{sc}] which provides an algorithmic procedure of formulating a field theory in the Riemann Cartan spacetime from the {\it corresponding} theory in the Minkowski spacetime, by localising the symmetry of the later under the Poincare group. We applied techniques similar to it in case of non-relativistic field theories. Of course there is fundamental difference between the structures of the Minkowski space time with the Galileo-Newton concept of euclidean space with universal time. The PGT localises the global Poincare symmetry of the parent theory where space and time were considered on equal footing according to  the special theory of relativity. Here, on the contrary, time has to be separated from space when devising the localisation prescription. According to Galilean concept time is not relative and thus the time translation parameter can only depend on time \cite{AHH, PP}. Space on the other hand is relative. Thus the spatial Galilean parameters on localisation are functions of both space and time. Our constructions, when geometrically interpreted, naturally leads to the Newton Cartan space-time \cite{BMM2}. Interestingly, with a vanishing time translation parameter the localisation procedure directly leads to spatially diffeomorphic theory. The advantage of the procedure is that the passage to Galilean symmetry in flat space is inbuilt. The entire approach is systematic without any ad-hoc assumptions.
 
  Gauging the Galileo symmetry from first principles is an intricate job. The Minkowski sapcetime is naturally a four dimensional manifold with non-degenerate metric that transforms as a second rank covariant tensor under Lorentz transformations. The physical fields constitute specific representations of the Lorentz group which include both spatial rotation and boosts. This facilitates the localisation of Poincare symmetry of a generic field. In the non-relativistic case there is no such luck. The Schrodinger field $\Psi({\bf{r},t})$ representing scalar particles transforms according to a projective representation \cite {JP, MC} whereas the gauge field transformation law under boost is not unique \cite{L}. In our previous work \cite{BMM1} only Schrodinger like scalar fields were considered. We started with a free theory with a generic lagrangian containing such fields only. The transformation of fields and its derivatives were worked out under global Galilean transformations. Naturally, the transformations of the temporal and spatial derivatives do not remain the same when the transformation parameters are localised. We introduced additional fields to construct {\it covariant derivatives} that transform under {\it local Galilean transformations} just as the {\it ordinary derivatives} do under {\it global Galilean transformations}. Another correction comes from the fact that the spatial Galilean transformations do not remain unimodular after localisation {\footnote{ Interestingly, the same fields which were introduced in converting the global covariant derivatives to local covariant derivatives are involved in the measure correction factor.}. This is the basic methodology which will be applied here.

   As was mentioned above, an important application of the formalism is in the theory of the fractional quantum Hall effect where the electrons move in two dimensional space under the action of a gauge field, the dynamics of which is dictated externally \cite{SW} or dynamically by the Chern-Simons term \cite{HS, S}. It is thus required to extend our formalism to include gauge fields in the field theory right at the beginning. This is all the more relevant as it has been reported \cite{HS, S} that the Chern Simons term poses problem in the formulation of non-relativistic diffeomorphism invariance. 
   
   In the present paper we generalize our earlier approach to include gauge fields. The nature of the Galilean concept of space-time makes this extension nontrivial. We assume that the theory invariant under global Galilean transformations contains a complex scalar field and an electromagnetic field. Usual first order theories are considered. Due to the presence of derivatives, these theories cease to be invariant under local Galilean transformations, i.e. when the transformation parameters are localised. In order to recover this invariance it is pertinent to realise that, after localisation, the transformations carry meaning as Galilean transformations with respect to local coordinates. Now, as already enunciated, the crucial point of our algorithm is to construct covariant derivatives that transform under local Galilean transformations as ordinary derivatives do under the corresponding global ones. For the construction of the covariant derivatives with respect to the local coordinates, we first define the covariant derivatives with respect to the global coordinates. The construction for the scalar field repeats the calculations already reported in \cite{BMM1} whereas new compensating fields are introduced corresponding to the gauge fields. Remarkable structural similarity of the global covariant derivatives is noticed. Also, necessity of treating the temporal and spatial components on different footing is observed. As to the conversion of the global to local covariant derivatives, identical mechanism works for all types of fields. The transformations of these new fields introduced in the second step are identical with those obtained in \cite{BMM1} which shows that these are connected with the geometry rather than with the specific fields. No wonder these are precisely those which were required to link the spacetime manifold with the Newton Cartan geometry
\cite{BMM2}.

 The formulation of  non-relativistic theories which will have diffeomorphism invariance in curved space is then discussed in full detail. This is achieved by a reinterpretation of the local Galilean symmetry as diffeomorphism symmetry. We start with vanishing time translation but keep the time dependence in the spatial transformation parameters. From the pool of the fields obtained in the localisation process we are able to construct a metric with the correct tensorial property in curved spacetime. In this geometric setting we view the transformation from local coordinates to global coordinates as transformation from the non-coordinate base to coordinate base which agree at the origin of the non-orthogonal coordinates. The appropriate transformation of the geometric objects such as scalars, vectors and other tensors are worked out. Note that though there is no time translation, the dependence of the spatial transformation parameters on time compels us to consider the time component of the geometric objects differently in the different bases. We work out the transformation rules of the covariant derivatives. The fall out is a step by step algorithm of introducing spatial diffeomorphism invariance. The passage to flat limit is manifest in our algorithm. 

We have also made a detailed investigation of the U(1) gauge symmetry. Contrary to the Galilean symmetry which was global to begin with, the gauge symmetry is already localized. The entire process of localization of the space-time symmetry eventually leading to a curved space interpretation, preserves this local U(1) gauge symmetry. We have explicitly demonstrated this for the two models analyzed here.

Our algorithm is then applied to definite problems which have appeared in the current literature on fractional quantum hall effect \cite{FQH}. Taking a complex scalar field interacting with the gauge field in flat space we localise it by the formalism derived here and formulate the theory in curved space time. To begin with, the gauge field is taken to be  external
and time independent diffeomorphism is considered. 
Applying our algorithm we construct the corresponding generally invariant theory in curved space. 
The resulting theory agrees well with that of \cite{SW} with a crucial difference; the gauge interaction gets modified due to the introduction of curvature.

It is the case of time dependent diffeomorphism where our theory predicts a completely new feature, namely the appearance of a new field. This is an auxiliary field that has no kinetic part. In this sense it can be considered as an external field acting on the electron which owes its existence to the curvature of space. 

 The challenging part is to include the dynamics of the gauge field. Specifically a crucial test is the inclusion of the CS dynamics as it has been reported \cite{HS, S} that spacetime diffeomorhism invariance is lost when CS dynamics is included. As one finds our systematic approach is equally applicable for the CS term. Thus we provide the complete formulation of the model of an electron moving in the curved space interacting with the CS gauge field. The formulation is such that at any stage of application the passage to the flat (euclidean) limit is inbuilt.

The paper is organized in six sections. In the following section we will discuss the general formalism in 2-space dimensions. Apart from a scalar, a gauge field is also considered, whose dynamics is kept open at this stage. At the end of the section we learn to modify a theory with global Galilean invariance to one with local Galilean invariance.
In section 3 we will present a novel way of converting the formalism to diffeomorphism in curved space. Applications of our formalism to two models including a comparison with existing results is provided in section 4. The models involve a Schrodinger field coupled, first, to an external vector field and next, to a vector field whose dynamics is governed by a Chern-Simons term. The issue of U(1)gauge symmetry is discussed in section 5. We have shown that the original gauge symmetry of the model is preserved in our localization process. The transformation of the complex scalar field and the gauge field have been worked out which are instrumental in demonstrating the local gauge inavriance in the backdrop of curved space. These ideas are applied to the two models considered here in section 4. We conclude in section 6.

\section{Gauging the Galilean symmetry of a model with scalar and vector fields}
We start with a theory given by the action
\begin{equation}
S = \int dx^0 d^2x {\cal{L}}\left(\phi_i, \partial_0 \phi_i, \partial_k \phi_i\right)\label{genaction}
\end{equation}
where  the index $0$ stands for time and $k = 1,2$ denote spatial coordinates. Often these will be represented collectively by  $\mu$.
The action (\ref{genaction}) is assumed to be invariant under the global Galilean transformation:
\begin{equation}
x^\mu\longrightarrow x^\mu+\xi^{\mu}\label{globalgalilean}
\end{equation}
where
\begin{equation}
\xi^{0}=-\epsilon,~~~~~~\xi^{i}=\epsilon^{i}+ \omega^{i}{}_{j}x^{j}-v^{i}x^0 
\end{equation}
 The time translation, space translation, spatial rotation and boost 
 parameters are constants,given by $\epsilon$, $\epsilon^{i}$, $\omega^{ij}$ and $v^{i}$
respectively.
The rotation parameter $\omega^{ij}$ are antisymmetric under interchange of the indices.
$\phi_i$ is a collection of fields which has definite transformation rules under 
(\ref{globalgalilean}) which leaves the action $S$ unchanged. The problem is to modify the theory (\ref{genaction}) so that the modified theory is invariant under the localised form of 
(\ref{globalgalilean}).
In \cite{BMM1} we have developed a systematic method of localisation including a complex scalar field in the action.
In this paper a vector field will be considered in addition. 

  A short review of our earlier work \cite{BMM1} will be appropriate at this stage. There we have considered a single scalar field only. When the Galilean transformations are localised, the transformation parameters  $\epsilon$, $\epsilon^{i}$, $\omega^{ij}$ and $v^{i}$ are no longer constants. The nature of non-relativistic spacetime dictates that the most general localisation of parameters is given by
\begin {equation}
\epsilon^0 \to \epsilon^0(x^0), \hskip .3cm\epsilon^{i}\to\epsilon^{i}(x^0,{\bf{r}}),\hskip .3cm \omega^{ij}\to  
\omega^{ij}(x^0,{\bf{r}}),\hskip .3cm v^{i}\to v^{i}(x^0,{\bf{r}})\label{localparameters}
\end{equation}
In order to give the local Galilean transformations a meaning we introduce local spatial coordinates $x^a, a =1,2$ which are trivially connected with the global coordinates $x^i$ by
\begin {equation}
x^a = \delta^a_i x^i\label{localcoordinates}
\end{equation}
The action  which was invariant under global Galilean transformations ceases to be so under the local version.
We demonstrated that the modified action
\begin{equation}
S = \int dx^0 d^2x \frac{M}{\theta}{\cal{L}}\left(\phi, \nabla_{\bar{0}}\phi, \nabla_a\phi\right)
\label{localactionold}
\end{equation}
 is invariant under the local Galilean transformations \footnote{The time component with respect to the local coordinates will be denoted by overbar. At this point there is no distinction between the time arrows perceived by the local and global observers. }. The quantities $\nabla_{\bar{0}}\phi$ and $\nabla_a\phi$ are covariant derivatives with respect to the local coordinates. They are related with the global covariant derivatives $D_0\phi$ and $D_k \phi$ by
\begin{eqnarray}
\nabla_{\bar{0}}\phi&=&\theta(\tilde{D}_0 \phi+\Psi^k \tilde{D}_k\phi)\nonumber\\
\nabla_a\phi&=&{\Sigma_a}^{k} \tilde{D}_k\phi
\label{finalcov}
\end{eqnarray}
where the global covariant derivatives $D_0\phi$ and $D_k \phi$ are defined as
\begin{eqnarray}
\tilde{D}_k\phi=\partial_k\phi+iB_k\phi\nonumber\\
\tilde{D}_0\phi=\partial_0\phi+iB_0\phi \label{firstcov}
\end{eqnarray}
The quantity $M$ in (\ref{localactionold}) is given by
\begin{equation}
 M = det{\Lambda_k}^a
\label{M}.
\end{equation}
where
${\Lambda_k}^a$ is the inverse of ${\Sigma_a}^k$
\begin{equation}
{\Lambda_k}^a{\Sigma_a}^l = \delta^l_k \hskip .15cm ;\hskip .15cm  {\Sigma_a}^k{\Lambda_k}^b = \delta^b_a 
\end{equation}
and 
$\theta$, $\Psi^k$, ${\Sigma_a}^k$, $B_0$ and $B_k$ are the new fields, the transformations of which have been worked out \cite{BMM1} so as to ensure the symmetry of (\ref{localactionold}) under the local Galilean transformations parametrised by (\ref{localparameters}). 

The procedure of getting (\ref{localactionold}) from (\ref{genaction}) can be understood from the following. From (\ref{genaction}) we can write the variation of the lagrangian under an arbitrary transformation $x^\mu \to x^\mu + \xi^\mu$ as
\begin{equation}
\Delta {{\cal{L}}} = \delta_0{{\cal{L}}} + \xi^{\mu}\partial_{\mu}{{\cal{L}}}+ \partial_{\mu}\xi^{\mu}{{\cal{L}}}\label{formvariation}
\end{equation}
 Here $\delta_0$ denotes the form variation given by 
\begin{equation}
\delta_0 \psi = \psi^{\prime}\left({\bf{r}}, x^0\right) - \psi\left({\bf{r}}, x^0\right)
\end{equation}
for any function $\psi\left({\bf{r}}, x^0\right)$.
For the global Galilean transformations $\partial_{\mu}\xi^{\mu} =0$. Also the fields and their derivatives transform in a way so that 
$$\delta_0{{\cal{L}}} + \xi^{\mu}\partial_{\mu}{{\cal{L}}} = 0$$
 For the local Galilean transformations the latter condition is satisfied when ordinary derivatives are replaced by the covariant derivatives.
But in this case $\partial_{\mu}\xi^{\mu} \neq 0$. The correction factor for the measure of the volume takes care of this and ensures that
 $\Delta {{\cal{L}}}= 0$. Naturally the action (\ref{localactionold}) is invariant.

Now we will use that same localization method to a more general case where the set of fields $\phi_i$ in (\ref{genaction}) contains a gauge field corresponding to electromagnetic interaction in addition to the scalar (Schrodinger) field. In other words, we consider the non-relativistic theory of complex scalar fields minimally interacting with vector gauge field in (2+1) dimensions, invariant under global Galilean transformations ({\ref{globalgalilean}}).  The action is  expressed as
\begin{equation}
S = \int dx^0 d^2 x {\cal{L}}\left(\phi,\partial_{\mu}{\phi}, A_\mu, \partial_\mu A_{\nu}\right)
\label{action}
\end{equation}
The action (\ref{action}) is known to be invariant under the local abelian gauge transformations
\begin{align}
\phi&\rightarrow\phi+i\Lambda\phi\nonumber\\A_{\mu}&\rightarrow A_{\mu}-\partial_{\mu}\Lambda
\label{gt}
\end{align} 
Apart from this invariance the action (\ref{action}) is invariant under the global Galilean transformations. We now discuss this issue in some details. Under the global Galilean transformations (\ref{globalgalilean}) the  complex scalar field $\phi$ transform as \cite{BMM1} 
\begin{eqnarray}
&\delta_0\phi = \epsilon\partial_0\phi - \eta^{i}\partial_i\phi + x^0v^{i}\partial_i\phi - imv^{i}x_i \phi\label{phi1}
\end{eqnarray}
where $\eta^i = \epsilon^i + \omega^{i}{}_{j}x^j $.
Consequently the derivatives vary as
\begin{eqnarray}
&\delta_0 \partial_k\phi=\epsilon\partial_{0}(\partial_{k}\phi)-\left(\eta^{i} -v^{i} x^0\right) \partial_{i}(\partial_{k}\phi)-
imv^i\partial_{k}(x_i\phi)
+\omega_k{}^{m}\partial_{m}\phi
\nonumber\\
&\delta_0 \partial_0\phi=\epsilon\partial_{0}(\partial_{0}\phi)-
(\eta^{i} - x^0v^i)\partial_{i}(\partial_{0}\phi)-imv^i x_i\partial_{0}\phi+v^{i}\partial_{i}\phi
\label{delkphi}
\end{eqnarray}
As we have mentioned earlier, due to the intricacies of the non-relativistic spacetime the transformation of various fields (under boost) must be determined
from case to case. The transformations of the gauge potential were obtained in 
\cite{L}. Of course $A_k$ transform as a vector under rotation while $A_0$ transform as a scalar under the same. Combining these the transformations of $A_\mu$ under global Galilean transformations are written as
 \begin{align}
\delta_0 A_0 &= \epsilon\partial_0 A_0 - \eta^{l}\partial_l A_0 + tv^{l}\partial_l A_0+v^l A_l\nonumber\\
\delta_0 A_i &= \epsilon\partial_0 A_i - \eta^{l}\partial_l A_i + tv^{l}\partial_l A_i+{\omega_i}^l A_l
\label{delA}
\end{align}
Then the transformations of their derivatives can be shown to be
\begin{align}
\delta_0 \partial_k A_0 &=\epsilon\partial_{0}(\partial_{k} A_0)-\left(\eta^{l} - x^0v^l\right)\partial_{l}(\partial_{k}A_0)+\omega_k{}^{l}\partial_{l}A_0+v^l\partial_k A_l\nonumber\\
\delta_0 \partial_0 A_0 &=\epsilon\partial_{0}(\partial_{0}A_0)-\left(\eta^{l} - x^0v^l\right) \partial_{l}(\partial_{0}A_0)+v^l \partial_l A_0+v^{l}\partial_{0}A_l
\label{delA0}
\end{align}
and 
\begin{align}
\delta_0 \partial_k A_i &=\epsilon\partial_{0}(\partial_{k} A_i)-\left(\eta^{l}-x^0v^l\right)\partial_{l}(\partial_{k}A_i) + {\omega_k}^l \partial_l A_i+\omega_i{}^{l}\partial_{k}A_l\nonumber\\
\delta_0 \partial_0 A_k &=\epsilon\partial_{0}(\partial_{0}A_k)-\left(\eta^{l}- x^0v^l\right)\partial_{l}(\partial_{0}A_k)+v^{l}\partial_{l}A_k
+\omega_k{}^{l}\partial_{0}A_l
\label{delAi}
\end{align}
These are the transformations that ensure 
\begin{equation}
\delta_0{{\cal{L}}} + \xi^{\mu}\partial_{\mu}{{\cal{L}}} = 0\label{reduced}
\end{equation}
Also here $\partial_\mu\xi^\mu = 0$. Together they keep $\delta S = 0$  under the global Galilean transformations,
where $S$ is given by (\ref{action}). 
 
Now we make the transformations local:
\begin{equation}
\xi^{0}=-\epsilon\left(x^0\right),~~~~~~\xi^i = \eta^i\left(x^0, {\bf{r}}\right) - v^i\left(x^0, {\bf{r}}\right)x^0
\label{localgalilean}
\end{equation}
where $\eta^i = \epsilon^{i}\left(x^0, {\bf{r}}\right)+\omega^{i}{}_{j}\left(x^0, 
{\bf{r}}\right)x^{j}$. Note the functional dependence of the various parameters of the local transformations. 
One should remember that after localisation these transformations can be viewed as Galilean transformations only locally. The final form of the local Galilean invariant theory will thus refer to the local coordinates. This explains the introduction of the local coordinates $x^a$ (see equation (\ref{localcoordinates})), 
notwithstanding the fact that in flat euclidean space they are trivially connected with the global coordinates. 

Once the parameters of the transformations are local
 the partial derivatives of $\phi, A_0, A_i$ with respect to space and time will no longer transform as (\ref{delkphi}, \ref{delA0}, \ref{delAi}). Following the gauge procedure one needs to introduce covariant derivatives which will transform covariantly as (\ref{delkphi}, \ref{delA0}, \ref{delAi}) with respect to the local coordinates. As we have shown in \cite{BMM1}, the first step in the process of localisation is to convert the ordinary derivatives into covariant derivatives with respect to the global coordinates. To begin with, introduce the gauge fields $B_\mu$ to define covariant derivatives of the complex scalar field $\phi$ with respect to space and time in global coordinate as,
\begin{eqnarray}
\tilde{D}_\mu\phi=\partial_\mu\phi+iB_\mu\phi
 \label{firstcovn}
\end{eqnarray}
Similarly new gauge fields $C_\mu, F_\mu$ will be introduced here to define the global covariant derivatives for  the fields $A_\mu$ as,
\begin{align}
\tilde{D}_\mu A_0 &=\partial_\mu A_0+iC_\mu A_0
\notag\\
\tilde{D}_\mu A_i &=\partial_\mu A_i+iF_\mu A_i
\label{firstcovg}
\end{align}
Note that different sets of gauge fields are introduced for $A_0$ and $A_i$, a typical signature of Galilean spacetime. Also note the structural similarity of the global covariant derivatives in each case.

In the next step the global covariant derivatives are converted to
the covariant derivatives with respect to space and time in local coordinates. For the complex scalar field these local covariant derivatives are defined as \cite{BMM1},
\begin{align}
\nabla_a\phi &={\Sigma_a}^{k}\tilde{D}_k\phi\notag\\
\nabla_{\bar{0}}\phi &=\theta(\tilde{D}_0 \phi+\Psi^k \tilde{D}_k\phi)
\label{nab}
\end{align}
where a is local index and k is global one,  introducing additional fields 
$\theta(x^0), \Psi^k(x^0, {\bf{r}}), {\Sigma_a}^k(x^0, {\bf{r}})$ in the process . We found that the local covariant derivative transform covariantly;
\begin{align}
 \delta_0 (\nabla_a\phi)= \epsilon \partial_0(\nabla_a\phi) -\left( \eta^i-x^0v^i\right) \partial_i(\nabla_a\phi) - imv^i\nabla_a\left(x_i\phi\right) + {\omega_a}^b\nabla_b\phi
  \label{covariantrule1}
\end{align}
provided the additional fields transform as
\begin{eqnarray}
{\delta}_0 B_{k} &=& \epsilon \dot{B}_k-{\partial}_k \left(\eta^i - x^0v^i\right) B_i  - \left(\eta^i-x^0v^i\right) {\partial}_i B_k  +  m{\partial}_kv^ix_i + m\left(v_k-{\Lambda_k}^a v_a\right)\nonumber\\
\delta_0 {\Sigma_a}^{k} &=&\epsilon{\dot{\Sigma}_a}^{k}+ {\Sigma_a}^{i}\partial_{i}
\left(\eta^{k}
-x^0v^{k}\right) - \left(\eta^{i} - x^0 v^{i}\right)\partial_{i}{\Sigma_a}^{k}+
{\omega_a}^b{\Sigma_b}^{k}
\label{delth1}
\end{eqnarray}
Here ${\Lambda_k}^a$ is the inverse of  ${\Sigma_a}^{k}$. For later convenience, we write the transformation of the inverse explicitly
\begin{equation}
\delta_0 \Lambda_{k}{}^{a}=\epsilon
{\dot\Lambda}_{k}{}^{a}- \Lambda_{l}{}^{a}\partial_{k}
\left(\eta^{l} -x^0v^l\right)
 - \left(\eta^{i}- x^0v^i\right)\partial_{i}\Lambda_{k}{}^{a}+
{\omega^a}_c\Lambda_{k}{}^{c}
\label{delLamb}
\end{equation}
Similarly to get the appropriate expression of $\delta_0(\nabla_{\bar{0}} \phi)$ as,
\begin{equation}
\delta_0 (\nabla_{\bar{0}}\phi)
=\epsilon\partial_{0}(\nabla_{\bar{0}}\phi)
-\left(\eta^{i}- x^0v^i\right)\partial_{i}(\nabla_{\bar{0}}\phi)-imv^i x_i\nabla_{\bar{0}}\phi
+v^{b}\nabla_{b}\phi
\label{covarianrule2}
\end{equation}
 we require that the introduced gauge fields should transform as,
\begin{align}
\delta_0 B_0 &=\epsilon \dot{B}_0+\dot{\epsilon}B_0-({\eta}^i-v^i x^0)\partial_i B_0-(\dot{\eta}^i-\dot{v}^i x^0)B_i
+v^i B_i +m\Psi^k{\Lambda_k}^a v_a+m{\dot{v}}^i x_i\notag\\
\delta_0\theta &=-\theta\dot{\epsilon}
+\epsilon\dot{\theta}\nonumber\\
\delta_0\Psi^k &=\epsilon{\dot{\Psi}}^k
+\dot{\epsilon}\Psi^k+\frac{\partial}{\partial x^0}({\eta}^k-x^0v^k)-(\eta^i-v^i x^0) \partial_i\Psi^k-x^0\Psi^i\partial_i v^k+\Psi^i\partial_i\eta^k+\frac{1}{\theta}v^b{\Sigma_b}^{k}
\label{delBt}
\end{align}
These transformations have already been reported in \cite{BMM1}. The new feature of the present model is the inclusion of the gauge fields $A_\mu$ in the original action.
We follow a similar procedure to construct the appropriate local covariant derivatives for these fields.
\begin{align}
\nabla_a A_{\bar{0}} &={\Sigma_a}^{k}\tilde{D}_k A_0\notag\\
\nabla_{\bar{0}} A_{\bar{0}} &=\theta(\tilde{D}_0 A_0+\Psi^k \tilde{D}_k A_0)\notag\\
\nabla_a A_b &=({\Sigma_a}^{k}\tilde{D}_k A_i){\delta^i}_b\notag\\
\nabla_{\bar{0}} A_b &=\theta(\tilde{D}_0 A_i+\Psi^k \tilde{D}_k A_i){\delta^i}_b
\label{nabA}
\end{align}
Plugging the expression of $\delta_0 {\Sigma_a}^{k}, \delta_0\Psi^k, \delta_0\theta$ the local covariant derivative will transform as the global one (see, equations (\ref{delA0}),(\ref{delAi})) 
\begin{align}
\delta_0(\nabla_a A_{\bar{0}}) &=\epsilon\partial_{0}(\nabla_{a} A_{\bar{0}})-\left(\eta^{l}-v^{l} x^0\right)\partial_{l}(\nabla_{a}A_{\bar{0}})
+\omega_a{}^{b}\nabla_{b}A_{\bar{0}}+v^b \nabla_a A_b\notag\\\delta_0 (\nabla_{\bar{0}} A_{\bar{0}}) &=\epsilon\partial_{0}(\nabla_{\bar{0}}A_{\bar{0}}) -\left(\eta^{l}-v^{l} x^0\right)\partial_{l}(\nabla_{\bar{0}}A_{\bar{0}})+v^b \nabla_b A_{\bar{0}}+v^{b}\nabla_{\bar{0}}A_b\notag\\
\delta_0 (\nabla_a A_b) &=\epsilon\partial_{0}(\nabla_{a} A_b)-\left(\eta^{l}- v^{l} x^0\right)\partial_{l}(\nabla_{a}A_b)+{\omega_a}^c \nabla_c A_b+\omega_b{}^{c}\nabla_{a}A_c\nonumber\\
\delta_0 (\nabla_{\bar{0}} A_b) &=\epsilon\partial_{0}(\nabla_{\bar{0}}A_b)-\left(\eta^{l}-v^{l} x^0\right)\partial_{l}(\nabla_{\bar{0}}A_b)
+v^{a}\nabla_{a}A_b
+\omega_b{}^{c}\nabla_{\bar{0}}A_c
\label{globalcov}
\end{align}
provided
\begin{align}
\delta_0 C_0 &=\epsilon\dot{C_0}
+\dot{\epsilon}C_0-(\dot{\eta}^i
-\dot{v}^{i}x^0)C_i-({\eta}^i-v^{i}x^0)\partial_i C_0+v^l C_l+i{A_0}^{-1}\dot{v}^l A_l
\nonumber\\
\delta_0 C_k&=\epsilon\dot{C}_k
-\partial_k({\eta}^i-v^{i}x^0)C_i-({\eta}^i-v^{i}x^0)\partial_i C_k+i{A_0}^{-1}\partial_k({v}^l) A_l\nonumber\\
\delta_0 F_0 &=\epsilon\dot{F}_0+\dot{\epsilon}F_0
-(\dot{\eta}^l-\dot{v}^{l}x^0)F_l-({\eta}^l-v^{l}x^0)\partial_l F_0+v^l F_l\notag\\
\delta_0 F_k &=\epsilon\dot{F}_k-\partial_k
({\eta}^l-v^{l}x^0)F_l-({\eta}^l-v^{l}x^0)\partial_l F_k
\end{align}
Certain interesting features in the construction of the local covariant derivatives for the gauge field are to be noted. First, we assume the same basic structure for constructing the corresponding global covariant derivatives as was done for the complex scalar field earlier. Second, it is remarkable that the same basic fields are employed to convert global to local covariant derivatives with the same set of transformation rules. This is why these fields are connected with the geometry of the non-relativistic spacetime \cite{BMM2}.

 The first stage of localization of Galilean transformation for the action is now over. Following the same approach stated in \cite{BMM1}, the action will be modified, replacing the    partial derivatives  by the local covariant derivatives. But, under the local Galilean transformation $\partial_\mu \xi^\mu \neq 0$ and a correction factor is required in the measure of integration (see equation (\ref{localactionold})). This prescription leads to the action
\begin{equation}
S = \int dx^0 d^2x \left(\frac{M}{\theta}\right){\cal{L}}\left(\phi,\nabla_\alpha{\phi},A_\alpha, 
\nabla_\alpha A_\beta\right)
\label{localaction}
\end{equation}
 where $\alpha,\beta \equiv {\bar{0}, a}$. The action (\ref{localaction})
  is invariant under the local Galilean transformations (\ref{localgalilean}).

Before closing this section let us emphasize the following points:
\begin{enumerate}
\item The theory (\ref{localaction}) is defined in flat (euclidean) background space.
\item The erection of the local coordinate system is to give meaning to the local Galilean transformations. Otherwise they are trivially connected with the global coordinates by (\ref{localcoordinates}). 
\end{enumerate}
In the following section we will find that the theory (\ref{localaction}) can be reinterpreted as a geometric theory where the connection between the global and the local coordinates will be nontrivial. This will lead naturally to a diffeomorphism invariant theory in space.

\section{Emergence of spatial diffeomorphism}
We will now show that our formalism leads to diffeomorphism invariant theory in 2-d space. Since the  goal is 2-d diffeomorphism in space we take the time translation in (\ref{localgalilean}) vanishing,
\begin {equation}
\epsilon(x^0)= 0\label{spacediff}
\end {equation} 
 The second equation of (\ref{delBt}) and (\ref{spacediff}) then show that  $\theta$ = constant. Without any loss of generality it can be taken to be one. 
 The local Galilean transformations with $\epsilon = 0$ is then equivalent to general coordinate transformation in space,
\begin{equation}
x^k \to x^k + \xi^k\label{genco}
\end{equation}
where $\xi^k$ is an arbitrary function of space and time defined in (\ref{localgalilean}). This indicates the possibility of reinterpreting the invariance of (\ref{localaction}) under (\ref{localgalilean})  as diffeomorphism invariance in curved space. But the theory (\ref{localaction})is formulated in terms of locally flat coordinates. When the background space is curved the local flat space is just the tangent space at the point of contact. In this new interpretation the coordinates labeled by $`a'$ define an orthogonal basis with origin at the point of contact. The coordinates labeled by $x$ define the coordinate basis in the curved space. In Cartan's formalism the connection between the two is established by the vielbeins. The fields ${\Sigma_a}^k$
can be reinterpreted as the vielbeins, as we will soon observe.

 Let us re-examine the structure of the transformation of ${\Sigma_a}^k$ which is obtained from (\ref{delth1}) under  the condition $\epsilon = 0$ as, 
 \begin{equation}
\delta_0 {\Sigma_a}^{k} = {\Sigma_a}^{i}\partial_{i}
\xi^k - \xi^i\partial_{i}{\Sigma_a}^{k}+
{\omega_a}^b{\Sigma_b}^{k}
\label{delth11}
\end{equation}
Note the dual aspects of the  transformation. With respect to the  coordinates $x^i$ it satisfies the transformation rules of a contravariant vector under the general coordinate transformation (\ref{genco}) whereas with respect to the coordinates $x^b$ it is a local rotation. From the transformation of $\Lambda_{k}{}^{a}$ given by (\ref{delLamb}) we find to our delight that it transforms as covariant vector under diffeomorphism (\ref{genco}) corresponding to its lower tier index $k$ while as an euclidean vector under rotation corresponding to its local index $a$. It will thus be reasonable to propose the following connection between local and global coordinates in the overlapping patch
\begin{equation}
dx_a = {\Sigma_a}^{k}dx_k\label{v}
\end{equation}
Note that, contrary to (\ref{localcoordinates}), the above connection has become nontrivial due to the geometric interpretation.

 

 We will next show that we can construct a metric (and its inverse) for the 2-d manifold from the fields ${\Sigma_a}^k$ and its inverse $\Lambda_{k}{}^{a}$. Let us define
\begin{equation}
g_{ij}=\delta_{cd}{\Lambda_i}^c {\Lambda_j}^d
\label{metric1}
\end{equation}
as a candidate for the metric.
From the transformation rules for ${\Lambda_i}^c$ we can prove that under the transformation (\ref{genco}) $g_{ij}$ transform as a covariant tensor 
\begin{equation}
\delta_0 g_{ij} =-\xi^k \partial_k g_{ij}-g_{ik}\partial_j\xi^k-g_{kj}\partial_i\xi^k
\label{diff} 
\end{equation}
The distance between two points is given by 
\begin{eqnarray}
dx_adx_a &=& {\Sigma_a}^{k}dx_k{\Sigma_a}^{l}dx_l
\nonumber\\
&=& \delta^{ab}{\Sigma_a}^{k}dx_k{\Sigma_b}^{l}dx_l
\nonumber\\
&=& g^{kl}dx_kdx_l
\end{eqnarray}
where, 
\begin{equation}
g^{kl}=\delta^{ab}{\Sigma_a}^{k}{\Sigma_b}^{l}
\label{invmetric}
\end{equation}
This $g^{ij}$ is the inverse of $g_{ij}$  and it transforms as a contravariant tensor. It can also be checked explicitly. Furthermore, $M = \rm{det}{\Lambda_i}^c = \sqrt{g}$, where $g$ is the determinant of $g_{ij}$.

The above developments suggest the following replacement
\begin{equation}
\int dx^0 d^2x \frac{M}{\theta} \to \int dx^0 d^2x \sqrt{g}\label{measure}
 \end{equation}
in (\ref{localaction}).
Note that this replacement is a transformation from local coordinates to global coordinates that charts {\it{2-dim curved space}}. By the reinterpretation of the fields we get curved geometry. The idea of spatial diffeomorphism that has surfaced in the theory of FQHE \cite{SW, HS} from an empirical point of view is thus shown to have deep connection with localisation of Galilean symmetry.

Now events happen not only in space but at a certain time instant also. Though we are working with vanishing time translation, the appearance of time in the diffeomorphism parameter $\xi$ makes
the time arrow relative at different points of curved space. The time component of the vectors in the local coordinate will not be simply equal with that of the curved space \footnote{That is why we have distinguished the corresponding indices from the beginning}. To relate the time components we will use the remaining field $\Psi^k$ and its transformation rule from (\ref{delBt}). Naturally this transformation rule does not show obvious geometric interpretation (spacetime is not a single manifold). However it fits with the emergent spatial diffeomorphism, as we will see.

From the practical point of view our theory gives a structural algorithm of constructing spatially diffeomorphic theory from the Galilean symmetric theories with the general structure of (\ref{action}). To establish this analogy we have to see how the transformations of the fields and the covariant derivatives obtained from the gauge approach in the previous section can be reinterpreted in the backdrop of curved space.

 The local coordinates map the tangent space at a space point. Geometric quantities are defined in the tangent space. Local 
coordinate basis is noncoordinate and orthogonal. They allow arbitrary rotations \footnote{The local system is tied to a point in the curved space. So Galilean boost is now no longer included in the local transformations. It is now absorbed in the spatial diffeomorphism}. We have the transformations of the physical fields 
$\phi$ and $A_{\bar{0}}, A_a$ at our disposal. Using equations (\ref{phi1} and (\ref{delA}) we can write these rules in the local coordinates as
\begin{eqnarray}
\delta_0\phi &=&  - \xi^{a}\partial_a\phi - imv^{a}x_a \phi\nonumber\\
\delta_0 A_{\bar{0}} &=& - \xi^{b}\partial_b A_{\bar{0}} +v^b A_b\nonumber\\
\delta_0 A_a &=&  - \xi^{b}\partial_b A_a +{\omega_a}^b A_b
\label{phi2}
\end{eqnarray}
In terms of these we will define the appropriate fields in the curved space.
Remember in this context that this mapping can only be achieved in the overlap of the two systems i.e in the neighborhood of origin of the local system.


We start with the scalar field $\phi$.
The transformation of the scalar field
in the curved space is obtained from (\ref{phi2}) as 
\begin{equation}
\delta_0\phi = - \xi^i\partial_i\phi\label{phic}
\end{equation}
Note that in the new interpretation the two descriptions match in the neighborhood of the origin of local coordinate system. This is why the last term of the corresponding equation of  (\ref{phi2})does not appear in (\ref{phic}).

Components of the vector field ${\bf{A}}$ are connected by a relation similar to  (\ref{v}),
\begin{equation}
A_a = {\Sigma_a}^{k}A_k
\label{Ac}
\end{equation}
The transformation of $A_a$ is the Galilean transformation given in (\ref{phi2})
and that of ${\Sigma_a}^{k}$ is given by (\ref{delth11}). The resulting transformation of $A_k$ in the curved basis is obtained by equating the form variations of both sides of (\ref{Ac}).
A straightforward calculation yields
\begin{equation}
\delta_0 A_k  =  - \xi^{i}\partial_i A_k - \partial_k\xi^iA_i\label{delAicurved}  
\end{equation}
These are the required ones for a covariant vector. 
In deriving (\ref{delAicurved}) we have used the following operator relation
\begin{align}
\xi_a\frac{\partial}{\partial_x^a} &=\xi_a \frac{\partial x_i}{\partial x_a}\frac{\partial}{\partial x_i}\notag\\&={\Sigma_a}^{k}\xi_k\Lambda_{i}{}^{a}\frac{\partial}{\partial x_i}\notag\\&=\xi_i\frac{\partial}{\partial_i}
\end{align}
which has been established using (\ref{v}).

Particular care is required for the time components of the fields. As has been already emphasized, though there is no time translation but time is involved in the spatial diffeomorphism parameters. The time component with respect to the local coordinates (denoted by an overbar on zero)
is to be related to the time component in curved coordinates by the following Ansatz
 \begin{equation}
 A_{\bar{0}}  = A_0 + \Psi^{k} A_k \label{delAbar0curved}  
\end{equation}
The transformation rule for $A_0$ is then worked out as 
\begin{equation}
\delta_0 A_0  =  - \xi^{i}\partial_i A_0 - \dot{\xi}^iA_i\label{delA0curved}  
\end{equation}
The structure of the above transformation is to be noted. The second term is dependent on the time variation of the diffeomorphism parameter  which can only be avoided if we consider time independent transformations. The structure of (\ref{delA0curved}) is the paradigm of the transformation of time components in the curved space, as will be subsequently observed.

After obtaining the transformations for the basic fields the geometric interpretation is established on firm ground. However, the issue of substituting
the covariant derivatives $\nabla_{{\bar{0}}}\phi$, $\nabla_k\phi$, $\nabla_aA_b$, $\nabla_{{\bar{0}}}A_a$, $\nabla_aA_{\bar{0}}$ and $\nabla_{{\bar{0}}}A_{\bar{0}}$ by appropriate derivatives with respect to the curved coordinates still remains. We denote these respectively by $D_0\phi$, $D_k\phi$, $D_kA_l$, $D_0A_l$, $D_kA_0$ and $D_0A_0$. The following definitions are proposed:
\begin{eqnarray}
\nabla_{a}\phi &=& {\Sigma_a}^k D_k\phi\nonumber\\
\nabla_{\bar{0}}\phi &=& D_0\phi + \Psi^kD_k\phi\nonumber\\
\nabla_{a}A_b &=& {\Sigma_a}^k {\Sigma_b}^l D_k A_l\nonumber\\
\nabla_{\bar{0}}A_a &=& {\Sigma_a}^{k}\left(D_0 A_k + \Psi^l D_l A_k\right)\nonumber\\
\nabla_a A_{\bar{0}} &=& {\Sigma_a}^{k}\left(D_k A_0 + \Psi^l D_k A_l\right)\nonumber\\
\nabla_{\bar{0}} A_{\bar{0}} &=& D_0A_0 +\Psi^k D_k A_0 +\Psi^k D_0 A_k +\Psi^k\Psi^l D_kA_l
\label{curvedcov}
\end{eqnarray}
Note that the construction of the time component of the covariant derivatives mimics our prescription (\ref{delAbar0curved}). 

The transformation laws of the new derivatives in curved space are once again obtained from the transformations rules (\ref{covariantrule1}), (\ref{covarianrule2}) and (\ref{globalcov}). To illustrate our method we take the transformation of $D_k\phi$ and show the calculation explicitly. Taking the form variation of both sides of the first equation of (\ref{curvedcov}) we get
\begin{equation}
\delta_0\left(\nabla_{a}\phi\right)=
\left(\delta_0{\Sigma_a}^k\right)D_k\phi
+ {\Sigma_a}^k\left(\delta_0D_k\phi\right)
\label{show}
\end{equation}
From (\ref{covariantrule1}) we write
\begin{equation}
\delta_0\left(\nabla_{a}\phi\right) =
-\xi^b\partial_b\left(\nabla_{a}
\phi\right) -imv^b\nabla_{a}\left(x_b\phi\right)+\omega_a{}^{b}\nabla_{b}\phi
\end{equation}
The last term of the above expression will have vanishing contribution because in the overlap of the two coordinate systems, $x_b\phi $ must be smoothly vanishing. Substituting this result on the left hand side of (\ref{show}) and using the transformation of $\Sigma_a{}^{k}$ we get the transformation $\delta_0D_k\phi$.
Working in an analogous way we get the transformation rules of the other curved space derivatives. The results are summarised as 
\begin{eqnarray}
\delta_0 D_k\phi &=& -\xi^i\partial_i\left(D_k\phi \right) - \partial_k\xi^i D_i\phi\nonumber\\
\delta_0 D_0\phi &=& -\xi^i\partial_i\left(D_0\phi \right) - \dot{\xi}^k D_k\phi\nonumber\\
\delta_0 D_k A_l &=& -\xi^i\partial_i\left(D_k A_l \right) - \partial_k\xi^m D_m A_l -\partial_l\xi^m D_k A_m\nonumber\\
\delta_0 D_0 A_k &=& -\xi^i\partial_i\left(D_0 A_k \right) - \partial_k\xi^l D_0 A_l -\dot{\xi}^l D_l A_k\nonumber\\
\delta_0 D_k A_0 &=& -\xi^i\partial_i\left(D_k A_0 \right) - \partial_k\xi^l D_l A_0 -\dot{\xi}^l D_k A_l\nonumber\\
\delta_0 D_0 A_0 &=& -\xi^i\partial_i\left(D_0 A_0 \right) - \dot{\xi}^k\left( D_k A_0 +  D_0 A_k\right)\label{varcurvedcov}
\end{eqnarray}
Note that all the curved space derivatives defined by (\ref{curvedcov}) transform canonically, following the transformations corresponding to their component labels established for the field components. For example, the expression for $\delta_0(D_k \phi)$ shows that $D_k\phi$ transforms as $A_k$ (see equation (\ref{delAicurved})). Similarly $D_0\phi$ transforms as $A_0$ (see (\ref{delA0curved})). The higher rank tensors like $D_k A_l$ transform appropriately.

For explicit calculations we will require expressions for the derivatives $D_k \phi, D_0\phi, D_k A_l, D_0 A_k, D_k A_0$ in terms of the basic fields with well defined transformations. These expressions are obtained by requiring consistency with (\ref{varcurvedcov}). Following this we define the derivatives $D_0\phi$ and $D_k\phi$ as,
\begin{align}
D_0\phi &= \partial_0\phi + i{\cal{B}}_0\phi\nonumber\\
D_k\phi &= \partial_k\phi + i{\cal{B}}_k\phi
\label{dkphi}
\end{align}
where the transformation rules for the fields $B_0$ and $B_k$ are given by,
\begin{align}
\delta_0 {\cal{B}}_0  &=  - \xi^{i}\partial_i {\cal{B}}_0 - \dot{\xi}^i {\cal{B}}_i\nonumber\\
\delta_0 {\cal{B}}_k  &=  - \xi^{i}\partial_i {\cal{B}}_k - \partial_k\xi^i {\cal{B}}_i\label{delBicurved}  
\end{align}
We observe that ${\cal{B}}_k$ transforms as a covariant spatial vector (see (\ref{delAicurved})) and ${\cal{B}}_0$ transforms in the same way as the time component of vectors are expected to transform in our formalism ( see equation (\ref{delA0curved}). This shows the internal consistency of our construction.

A word about the introduction of the new field ${\cal{B}}$ is useful. Observe that the set of vector fields A were present in the original model. The new vector fields ${\cal{B}}$ emerge from the localization prescription that leads to our formulation in curved space.

Similarly we define the other derivatives acting on `A's in the following way,
\begin{eqnarray}
 D_i A_k &=& \left(\partial_i A_k - \partial_k A_i\right) + i({\cal{B}}_i A_k-{\cal{B}}_k A_i)\nonumber\\
 D_0 A_k &=& \left(\partial_0 A_k - \partial_k A_0\right) + i({\cal{B}}_0 A_k-{\cal{B}}_k A_0\nonumber\\
 D_k A_0 &=& \left(\partial_k A_0 - \partial_0 A_k\right) + i({\cal{B}}_k A_0-{\cal{B}}_0 A_k)
\label{covexpand}
\end{eqnarray}
such that they satisfy the transformation rules (\ref{varcurvedcov}).

The algorithm for the construction of the spatially diffeomorphic theories can now be summarised:
\begin{enumerate}
\item Start from a non relativistic Galilean invariant theory.
\item Gauge the Galilean symmetry by replacing the derivatives of the field by the corresponding local covariant derivatives.
Also correct the measure appropriately as in (\ref{localaction}). The resulting theory is now locally Galilean invariant theory.
\item Take time translation vanishing. The local Galilean transformations are then equivalent to general coordinate transformations in curved space.  
\item Formulate the theory as a theory invariant under general coordinate transformations in a curved space by the substitution (\ref{measure}) and by replacements of the covariant derivatives in the action (\ref{localaction}) by the covariant derivatives in the curved space. Use the definitions (\ref{curvedcov}).

\item The diffeomorphic theory obtained in the above procedure will contain the fields $\Sigma_{a}{}^{k}$ and $\Psi^k$. The fields $\Sigma_{a}{}^{k}$ will be grouped to give rise to tensors in the curved space e.g the metric tensor. The fields $\Psi^k$ are independent fields in the theory without any kinetic term.
\end{enumerate}

\section{Applications and comparison with existing results}
In this section we will discuss a couple of applications of our general formalism and make a comparison with existing results. The first model will be a complex Schrodinger field theory in the presence of an external vector field. The other model to be considered will involve a vector field whose dynamics is generated by a Chern-Simons term.\\ 
\subsection{Complex Schrodinger field theory in the presence of an external vector field}

As we have mentioned in the introduction the most important application of spatial diffeomorphism is in the theory of fractional quantum Hall effect  \cite{SW}. 
It will thus be useful to start from the example which models a non relativistic  electron moving in an external gauge field
given by the action 
\begin{equation}
S = \int dx^0  \int d^2x_k  \left[ \frac{i}{2}\left( \phi^{*}\Delta_{0}\phi-\phi \Delta_0\phi^{*}\right) -\frac{1}{2m}\Delta_k\phi^{*}\Delta_k\phi\right]
\label{globalaction2} 
\end{equation} 
where
\begin{eqnarray}
\Delta_{0}\phi = \partial_0\phi + iA_0\phi\nonumber\\
\Delta_{k}\phi = \partial_k\phi + iA_k\phi
\end{eqnarray}
and $A_\mu$ is the external gauge field. The theory (\ref{globalaction2}) is invariant under global Galilean transformations (\ref{globalgalilean}) as can be checked explicitly. 

Simplifying (\ref{globalaction2}) we can get,
\begin{align}
S = \int dx^0 \int d^2 x_k &\left[ \frac{i}{2}\left( \phi^{*}\partial_{0}\phi-\phi\partial_0\phi^{*}\right)-\phi^{*}\phi A_0-\frac{1}{2m}\partial_k\phi^{*}\partial_k\phi-\right.\notag\\
& \left.\frac{{A_k}^2}{2m}\phi^{*}\phi+\frac{i}{2m}A_k(\phi^{*}\partial_{k}\phi-\phi\partial_k\phi^{*})
\right]  
\end{align}

 The corresponding theory invariant under local Galilean transformations (\ref{localgalilean}), according to our algorithm,
is
\begin{align}
S = \int dx^{\bar{0}} \int d^2x_a \frac{M}{\theta}&\left[\frac{i}{2}\left(\phi^{*}\nabla_{\bar{0}}\phi-\phi \nabla_{\bar{0}}\phi^{*}\right)-\frac{1}{2m}\nabla_a\phi^{*}\nabla_a\phi -\phi^{*}\phi A_{\bar{0}}-\right.\notag\\&\left.\frac{{A_a}^2}{2m}\phi^{*}\phi+\frac{i}{2m}A_a(\phi^{*}\nabla_{a}\phi-\phi\nabla_a\phi^{*})
\right]
\label{localscintaction} 
\end{align}
In the following we will consider spatial diffeomorphism ($\epsilon = 0$) where $\theta = ~1$.
We can then transform our results in a geometric setting following the algorithm given at the end of section 3.
 
Let us first consider the special case when $\xi$, the spatial diffeomorphism parameter, is time independent. The third equation of (\ref{delBt}) shows that, along with the time independence of $\xi$, $\Psi_k = 0$ may be chosen. Under this condition,  $\nabla_{\bar{0}}\phi=D_0\phi$. After some algebra the action (\ref{localscintaction}) reduces to, 
 \begin{align*}
S &= \int dx^0 \int d^2x (det{\Lambda_k}^a)\left[\frac{i}{2}\left(\phi^{*}D_{0}\phi-\phi D_0\phi^{*}\right)-\phi^{*}\phi A_0-{\Sigma_a}^k{\Sigma_a}^l \left(\frac{1}{2m}D_k\phi^{*}D_l\phi\right) -\right.\notag\\&\left.{\Sigma_a}^k{\Sigma_a}^l\left(\frac{1}{2m}A_k A_l\phi^{*}\phi\right)+{\Sigma_a}^k{\Sigma_a}^l\left(\frac{i}{2m}A_k(\phi^{*}D_{l}\phi-\phi D_l\phi^{*})\right)
\right]
\end{align*}
Using the definition of metric (\ref{invmetric}) this is reduced to a generally covariant theory in the curved space
\begin{align}
S &=\int dx^0 d^2x (det{\Lambda_k}^a)\left[ \frac{i}{2}\left( \phi^{*}(D_{0}+iA_0)\phi-\phi(D_0-iA_0)\phi^{*})\right)\right.\notag\\&\left. -g^{kl}\frac{1}{2m}(D_k-iA_k
)\phi^{*}(D_l+iA_l)\phi\right]
\label{localscaction1} 
\end{align}
The action (\ref{localscaction1}) can now be written as a non-relativistic diffeomorphism invariant action,
\begin{equation}
S = \int dx^0 d^2x \sqrt{g}\left[ \frac{i}{2}\left( \phi^{*}{\bar{D}}_{0}\phi-\phi{\bar{D}}_0\phi^{*}\right) -g^{kl}\frac{1}{2m}\bar{D}_k\phi^{*}\bar{D}_l
\phi\right]\label{diffaction3}
\end{equation}
where 
\begin{eqnarray}
\bar{D}_{0}\phi = D_0 \phi+i A_0\phi=\partial_0\phi +i\left(A_0 + {\cal{B}}_0\right)\phi\nonumber\\
\bar{D}_{k}\phi = D_k\phi+iA_k\phi=\partial_k\phi +i\left(A_k + {\cal{B}}_k\right)\phi
\label{dbar}
\end{eqnarray}

So we can interpret from the result that localization of Galilean symmetry for the non-relativistic field theoretic model of complex scalar fields interacting with vector field in flat space gives a theory with an action invariant under general coordinate  transformation in curved space. Note that we have considered the spatial diffeomorphism parameter as time independent and there is no time translation.

At this point we can compare our results with that of \cite{SW}. They obtained spatial diffeomorphism by following the minimal coupling prescription as,
\begin{equation}\label{free-L}
  S = \int dt dx \sqrt{g}\left[\frac{i}{2} (\psi^{\dagger}\partial_t\psi-\psi\partial_t\psi^\dagger)
  - A_0\psi^{\dagger}\psi 
  - \frac{g^{ij}}{2m}(\partial_i\psi^{\dagger}-iA_i\psi^{\dagger})(\partial_j\psi+iA_j\psi)\right].
\end{equation}
which is invariant under infinitesimal transformations,
\begin{align}\label{3d-gci}
  x^i \to x^{i'} &= x^{i'}(x^i), \quad
  \psi(t,x) \to \psi(t,x') = \psi(t,x),\nonumber\\ 
  A_0(t,x)\to A_0'(t,x') &= A_0 (t,x), \quad
A_i(t,x) \to A_{i'} (t,x') = \frac{\partial x^i}{\partial x^{i'}}
    A_i(t,x)\nonumber\\  g_{ij}(t,x) \to g_{i'j'}(t,x') &= \frac{\partial x^i}{\partial x^{i'}}
    \frac{\partial x^j}{\partial x^{j'}} g_{ij}(t,x).
\end{align}
 when the fields transform as {\footnote{Note that, to make a comparison, we have set the gauge parameter in \cite{SW} to zero, since we consider only diffeomorphism symmetry.}},
\begin{align}\label{static-gci}
  \delta\psi &=- \xi^k\partial_k\psi, 
 \quad \delta A_0 = -  \xi^k\partial_k A_0,\nonumber\\
  \delta g_{ij} &= -\xi^k \partial_k g_{ij} - g_{ik}\partial_j \xi^k -
     g_{kj}\partial_i \xi^k, \quad
  \delta A_i =- \xi^k\partial_k A_i - A_k \partial_i\xi^k .
  \end{align}
 The action (\ref{free-L}) agrees with (\ref{diffaction3}) with the proviso that `A' is replaced by `A+B'. In the time independent case the transformations of basic fields given above becomes identical with that obtained here in (\ref{phic}, \ref{delA0curved}, \ref{delAicurved}, \ref{delBicurved}).
 
When the diffeomorphism parameter $\xi^i$ is time dependent the real difference comes up. Now $\Psi^k=0$ is not admissible. Then the diffeomorphism invariant action in the curved space becomes 
\begin{eqnarray}
S &=& \int dx^0 d^2x \sqrt{g}[ \frac{i}{2}\left(\phi^{*}{\bar{D}}_{0}\phi-\phi{\bar{D}}_0\phi^{*}\right) -g^{kl}\frac{1}{2m}\bar{D}_k\phi^{*}\bar{D}_l\phi
\nonumber\\&+& \frac{i}{2}\Psi^k\left(\phi^{*}{\bar{D}}_{k}\phi
-\phi{\bar{D}}_k\phi^{*}\right)]
\label{diffaction12}
\end{eqnarray}
Note that we do not demand any special transformation for the time dependent case. Identical transformation laws for the basic fields ensure the invariance of the action (\ref{diffaction12}). This is to be contrasted with \cite{SW} where the same action is retained but the transformation rules of the basic fields change in a non-canonical way {\footnote{These are given in equation (17) of \cite{SW}}}. This is not surprising because the results of \cite{SW} are obtained in an adhoc manner, based on `trial and error' method as the authors of \cite{SW} admitted. On the other hand our analysis does not distinguish between time dependent and time independent cases, both of which can be obtained in holistic manner following our localization procedure.

Before finishing this comparison we would like to draw attention to a crucial point. In the general case when $\xi^i$ is time dependent a set of non-canonical transformations of the fields is given in \cite{SW} where the gauge transformations also contribute. To derive the flat limit  of these transformations they put as usual $g_{ij}=\delta_{ij}$. The surprising thing is that in the flat limit the Galilean transformation can only be recovered if one assumes a particular correlation between the gauge parameter and the boost parameter. This can hardly be motivated on any fundamental premises. Also observe that the passage to flat limit is naturally inbuilt in our construction. Thus there is no trouble in recovering Galilean invariance. It is just required to replace the covariant derivative by the ordinary derivative and the metric by $\delta_{ij}$. A simple inspection of (\ref{diffaction12}) and (\ref{globalaction2}) confirms the above. \footnote{Note that $\Psi^k$ vanishes when the covariant derivative is replaced by the ordinary derivative.}
\subsection{Inclusion of the Chern-Simons term in the action}
Another landmark problem is the inclusion of the Chern-Simons (CS) term in the action \cite{HS, S}. The CS action is given by 
\begin{equation}
S_{CS} = \int d^3x \frac{\kappa}{2}\epsilon^{\mu\nu\lambda}A_\mu\partial_\nu A_\lambda
\end{equation}
and can be coupled with both relativistic and non-relativistic models \cite{D}. It will be convenient to break the action in spatial and temporal parts,
\begin{equation}
S_{CS} = \int dx^0  \int d^2x_k \frac{\kappa}{2}\epsilon^{ij}\left(A_0\partial_i A_j-A_i\partial_0 A_j+A_i\partial_j A_0\right)
\label{globalactioncs} 
\end{equation} 

 It can be shown that (\ref{globalactioncs}) is invariant under the global Galilean transformation using the variations (\ref{delA}). Following the method to localize the Galilean transformation stated in previous section, we can get the corresponding action invariant under the the local Galilean transformations as
\begin{align}
S = &\int dx^{\bar{0}} \int d^2x_a \frac{M}{\theta}
\frac{\kappa}{2}\epsilon^{ab}\left(A_{\bar{0}}\nabla_a A_b-A_a\nabla_{\bar{0}} A_b+A_a\nabla_b A_{\bar{0}}\right)
\label{localscaction} 
\end{align}
By our construction
this action (\ref{localscaction}) is invariant under (\ref{localgalilean}). This can also be checked explicitly.

 Now our algorithm given above in section 3 allows us to construct the spatially diffeomorphic action as follows: 
\begin{eqnarray}
S &=& \int dx^0 d^2x \sqrt{g}\frac{\kappa}{2}\epsilon^{ab}{\Sigma_a}^k{\Sigma_b}^l\left[\left(A_0D_k A_l-A_kD_0 A_l+
A_kD_lA_0\right)\right.\nonumber\\ 
&+&\left.\Psi^m A_m D_k A_l+\Psi^m A_k\left(D_l A_m - D_m A_l\right)\right]
\label{curvedaction} 
\end{eqnarray}
Note that $\epsilon^{ab}$ is a tensor under local (orthogonal) transformations. Thus
\begin{equation}
{\Sigma_a}^k{\Sigma_b}^l\epsilon^{ab} ={\tilde{\epsilon}^{kl}} 
\end{equation}
where $\tilde{\epsilon}^{kl}$ is the Levi Civita tensor in the curved space. It is related to the numerical tensor 
$\epsilon^{kl}$ by,
\begin{equation}
\tilde{\epsilon}^{kl}=\frac{1}{\sqrt g}\epsilon^{kl}
\end{equation}

Then the CS action in curved space is obtained from the above equations as,
\begin{eqnarray}
S &=& \int dx^0 d^2x \frac{\kappa}{2}{{\epsilon}^{kl}}\left[\left(A_0D_k A_l-A_kD_0 A_l+
A_kD_lA_0\right)\right.\nonumber\\ 
&+&\left.\Psi^m A_m D_k A_l+\Psi^m A_k\left(D_l A_m - D_m A_l\right)\right]\label{curvedaction1}
\end{eqnarray}
Now the derivatives $D_{\mu}A_{\nu}$ are substituted from (\ref{covexpand}).
\begin{align}
S &=\int dx^0 d^2x \frac{\kappa}{2}{{\epsilon}^{kl}}\left[\left(A_0 (\partial_k A_l-\partial_l A_k+iB_kA_l-iB_lA_k)-A_k(\partial_0 A_l-\partial_l A_0+i B_0 A_l-iB_lA_0)\right. \right.\nonumber\\  & \left. \left.  +
A_k(\partial_l A_0-\partial_0 A_l+iB_l A_0-iB_0 A_l)\right) +
\Psi^m[ A_m (\partial_k A_l-\partial_l A_k+iB_kA_l-iB_lA_k)\right.\notag\\& \left.+ A_k(\partial_l A_m-\partial_m A_l+iB_l A_m-iB_mA_l)- A_k(\partial_m A_l-\partial_l A_m+iB_m A_l-iB_lA_m)\right]]\label{gabbar}
\end{align}
Exploiting the antisymmetric property of ${\epsilon}^{kl}$, (\ref{gabbar}) further reduces to,
\begin{align}
S &=\int dx^0 d^2x \frac{\kappa}{2}{{\epsilon}^{kl}}\left[2\left(A_0\partial_k A_l-A_k\partial_0 A_l+
A_k\partial_lA_0\right)\right.\nonumber\\ 
&+\left.2\Psi^m[ A_m \partial_k A_l+ A_k(\partial_l A_m-\partial_m A_l)\right]]\notag\\&=\int dx^0 d^2 x \kappa 
[\epsilon^{\mu\nu\lambda} A_{\mu}\partial_{\nu}A_{\lambda}+\Psi^m \epsilon^{kl}[ A_m \partial_k A_l+ A_k(\partial_l A_m-\partial_m A_l)]]
\label{gabbar1}
\end{align}
Note that the ${\cal{B}}$ field has dropped out from the above expression. Effectively, therefore, the Chern-Simons interaction receives a correction to its original form.

It may be shown that the above action, under the general coordinate transformations (\ref{delAicurved}), (\ref{delA0curved}) and (\ref{varcurvedcov}), changes as
\begin{equation}
\delta S =\int dx^0 d^2x \kappa \partial_i\left[\xi^i\epsilon^{kl}\left(A_0\partial_kA_l - A_k\partial_0A_l + A_k\partial_lA_0\right)\right]
\end{equation}
The integrand is a total derivative and drops to zero when integrated over space. This proves that the action is invariant under the general coordinate transformations.

The Chern - Simons action has proved to be very useful in the study of fractional quantum Hall effect. In this context it may be noted that the Chern - Simons action is reported \cite{HS} to break the diffeomorphism symmetry. This has been a major obstacle in applying theories with Chern - Simons term in curved space. To recover the lost invariance it is essential to introduce correction terms. In our opinion these features are manifestations of the ad hoc prescription used to achieve non relativistic diffeomorphism invariance from a theory defined in flat space. Our approach on the other hand naturally leads to an appropriate Chern - Simons theory in curved space, without any adhoc assumptions or corrections.

\section{Comments on U(1) gauge symmetry}
In this section we will analyze the issue of `gauge invariance' in our theory in more details. First, we will discuss the gauge invariance of the localized Galileo symmetric model given in (\ref{localaction}). When the Galilean symmetry is global the gauge transformations are given by (\ref{gt}) in which case the combination $(\partial_{\mu}\phi+iA_{\mu}\phi)$ transforms covariantly as follows,
\begin{equation}
\partial_{\mu}\phi+iA_{\mu}\phi\rightarrow (1+i\Lambda)(\partial_{\mu}\phi+iA_{\mu}\phi)
\end{equation}
 When the Galilean symmetry is localized the partial derivatives $\partial_{\mu}\phi$ are replaced by $\nabla_a \phi$. Now the combination $(\nabla_a\phi+iA_a\phi)$ transforms covariantly as,
 \begin{equation}
\nabla_{a}\phi+iA_{a}\phi\rightarrow (1+i\Lambda)(\nabla_{a}\phi+iA_{a}\phi)
\end{equation}
This is achieved for the following transformations of the basic fields ,
\begin{align}
\phi &\rightarrow\phi+i\Lambda\phi\notag\\A_a \rightarrow A_a-\nabla_a \Lambda,~~& A_{\bar{0}} \rightarrow A_{\bar{0}}-\nabla_{\bar{0}} \Lambda
\end{align}
where,
\begin{equation}
\nabla_a\Lambda=\Sigma_a{}^{k}\partial_k\Lambda,~~\nabla_{\bar{0}}\Lambda=\partial_0 \Lambda+\Psi^m\partial_m \Lambda
\label{gl}
\end{equation}
From (\ref{Ac}, \ref{delAbar0curved}) and (\ref{gl}) we can analyze the behavior of the external gauge field in curved space under the gauge transformation. It is given by
\begin{equation}
A_k\rightarrow A_k-\partial_k\Lambda,~~A_0\rightarrow A_0- \partial_0\Lambda
\label{gc}
\end{equation}
and has the expected form suggested by (\ref{gt}). Now we will discuss the gauge invariance of two different cases in section 4 explicitly.
\subsection{Gauge invariance for complex Schrodinger field theory in the presence of an external vector field}
An explicit demonstration of the gauge invariance of the action (\ref{diffaction12}) is straightforward. Let us first consider the structure of the derivatives appearing in (\ref{dbar}). Then under the gauge transformation (\ref{gl}, \ref{gc}) it is easy to show that these derivatives transform covariantly.
\begin{equation}
\bar{D}_{0}\phi\rightarrow (1+i\Lambda)\bar{D}_{0}\phi,~~~\bar{D}_{k}\phi\rightarrow (1+i\Lambda)\bar{D}_{k}\phi
\label{cog}
\end{equation}
Note that the new fields $(\cal{B})$  do not transform under the gauge transformation. Indeed if ${\cal{B}}$ changes under gauge transformation then the above covariant property is lost. The point is that introduction of ${\cal{B}}$ was a consequence of localization of spacetime symmetry. So ${\cal{B}}$ changes under the general coordinate transformation but not under the gauge transformation. It may be recalled that the original gauge symmetry of the model is already localized ( See for instance the discussion below {\ref{action}). 

Using the covariant property of the derivatives (\ref{cog}) it is easy to show that the action (\ref{diffaction12}) is invariant under the gauge transformation.

\subsection{Gauge invariance in Chern-Simons interaction}
 Under the gauge transformation (\ref{gc}) the action (\ref{gabbar1}) can be shown to be invariant. The first piece is identically the Chern-Simons term whose gauge invariance is well known. The terms in the second parenthesis give a correction to the Chern-Simons action which will vary under the gauge transformation as,
\begin{align}
\delta {\cal{L}}&=2\Psi^m \epsilon^{kl}[(\partial_m\Lambda)(\partial_k A_l)+(\partial_k \Lambda)(\partial_l A_m-\partial_mA_l)\notag\\&=2\epsilon^{kl}[\partial_m(\Psi^m \Lambda \partial_k A_l)+\partial_k(\Psi^m \Lambda(\partial_l A_m-\partial_m A_l))\notag\\&-\Lambda[(\partial_m\Psi^m)(\partial_k A_l)+(\partial_k \Psi^m)(\partial_l A_m-\partial_m A_l)]]
\end{align}
The second term proportional to $\Lambda$ vanishes identically. Thus $\delta{\cal{L}}$ is a pure boundary so that the action (\ref{gabbar1}) remains invariant.

Note that $\Psi^m$ which appears in the above example is actually related to the Newton-Cartan data as was discussed in our earlier work \cite{BMM2}. 
\section{Conclusion}
The problem of formulating a Galilean invariant theory in euclidean space and universal time into a diffeomorphism invariant theory in curved space has been addressed in the paper. We have considered a generic theory containing a Schrodinger field and a gauge field. A complete algorithm was given and its applications were discussed in relation to the model of an electron moving in  two dimensional space under the action of an external electromagnetic field as well as under a field whose dynamics was dictated by the Chern-Simons (CS) term. The flat (euclidean) limit was reproduced naturally without any assumptions.

The algorithm given in this paper can be divided in two steps. In the first step a theory invariant under the global Galilean transformations was taken. The symmetry was localised following the general notions adopted for constructing Poincare gauge theory \cite{U} - \cite{sc},  modulo nontrivial modifications due to the difference in the concept of time occurring in relativistic and non-relativistic theories.The fundamental difference between the Minkowski space time with Galilean space and universal time makes the problem highly intricate. The localisation process naturally separated time from space. Local coordinates had to be assumed to give local Galilean transformations a meaning notwithstanding the fact that at the flat (euclidean) stage their relation with the global coordinates was trivial. Ordinary derivatives were replaced by covariant derivatives with respect to local coordinates by introducing new fields. Also the measure of the integration was corrected appropriately. This resulted in a theory in local coordinates invariant under local Galilean transformation. 
 
 Several new fields were introduced in the first step. These new `gauge' fields can be divided in two classes. In the first category we have a group of fields  which are similar for all kinds of parent fields. These fields were associated with geometry. The new fields in the other class were specific to the fields of the theory.

In the second step the resulting theory was geometrically interpreted. The geometric content of the construction was then studied using the first category of fields. We reinterpreted the global coordinates as coordinates charting the curved space whereas the local coordinates were identified with locally euclidean coordinates. A spatial metric was constructed with all the desired properties and the transformations of the various fields were worked out. The geometric interpretation was thus firmly established. An algorithm with step by step instructions was formulated to derive the diffeomorphic theory in the curved space. 

The algorithm derived in the paper was then applied to the very important problem of an electron moving in 2-d space under an external field. The similarities and points of departure of our results with those obtained in \cite{SW, HS, S} were emphasized. We then took an electromagnetic field whose dynamics was dictated by the Chern-Simons term. No problem was encountered in writing the corresponding generally covariant theory in space. This may be compared with other approaches where covariantisation of the C-S term poses problems.

As a final remark we note that the issue of U(1) gauge symmetry was also discussed in some details. The relevant derivatives that appeared after the localization process were shown to transform covariantly under this gauge transformation. This was instrumental in proving the gauge invariance of the model discussed here, particularly in the example of Schrodinger field coupled with an external field. For the Chern- Simons theory the additional fields introduced during localization procedure dropped out. As happens for C-S theory the gauge variation changed the lagrangian by a total derivative so that the action remained invariant.

\end{document}